\documentclass[prl, twocolumn, floatfix]{revtex4}
\setcitestyle{super}

\usepackage{graphicx}
\usepackage{hyperref}
\usepackage{amsmath}
\usepackage{color}

\begin{document}

\title{Magnetic Phase Diagram of Light-mediated Spin Structuring in Cold Atoms}

\author{G.\ Labeyrie$^{1}$\footnote{To whom correspondence should be addressed.}, I.\ Kre\v si\' c$^{2,3}$, G.~R.~M. Robb$^{2}$, G.-L. Oppo$^{2}$, R.\ Kaiser$^{1}$, and T.\ Ackemann$^{2}$}
\affiliation{$^{1}$Universit\'{e} C\^{o}te d'Azur, CNRS, Institut de Physique de Nice, 06560 Valbonne, France}
\affiliation{$^{2}$SUPA and Department of Physics, University of Strathclyde,Glasgow G4 0NG, Scotland, UK}
\affiliation{$^{3}$Institute of Physics, Bijeni\v cka cesta 46, 10000, Zagreb, Croatia}

\begin{abstract}
When applying a red-detuned retro-reflected laser beam to a large cloud of cold atoms, we observe the spontaneous formation of 2D structures in the transverse plane corresponding to high contrast spatial modulations of both light field and atomic spins. By applying a weak magnetic field, we explore the rich resulting phase space and identify specific phases associated with both dipolar and quadrupolar terms of the atomic magnetic moment. In particular we demonstrate spontaneous structures in optically induced ground state coherences representing magnetic quadrupoles.
\end{abstract}

\maketitle

\section{Introduction}

Since the first observation of the spontaneous appearance of hexagonal structures in the transverse cross-section of laser beams counterpropagating in sodium vapor~\cite{grynberg88b}, self-organized optical structures have been investigated in  various interaction geometries (cavities~\cite{lugiato87,kreuzer91,geddes94b,scroggie94,tlidi94,oppo94,kuszelewicz00}, single-mirror feedback arrangements~\cite{firth90,dalessandro91,macdonald92a,honda93,ciaramella93,grynberg94,ackemann94,aumann97,schwab99}, counterpropogating beams~\cite{grynberg88,pender90,firth90,gauthier90,petrossian92,dawes05}, hybrid systems such as liquid crystal light valves~\cite{akhmanov88,pampaloni93,thuering93}) and many different nonlinear materials (liquid crystals~\cite{macdonald92a,ciaramella93}, alkaline atoms~\cite{grynberg88,grynberg94,ackemann94,dawes05}, semiconductors~\cite{kuszelewicz00}, photorefractives~\cite{honda93,schwab99}). These structures arise from the interaction of diffraction providing spatial coupling and optical nonlinearities. Analysis was mainly based on optical wave mixing mediated by the medium (e.g.\ explicitly in the class A limit in cavity transverse nonlinear optics~\cite{lugiato87,geddes94b,scroggie94} and in most treatments of counterpropagating beams~\cite{grynberg88,pender90,firth90}), although obviously there is always a corresponding structure in the matter variable the optical field is coupling to. In particular, it was known that the optical structures created via optical pumping nonlinearities in alkaline vapors will create a corresponding magnetization in the atomic ground states, which, under the experimental conditions used  could be analysed in the framework of a spin-1/2 system  and the resulting structures corresponded to modulations of magnetic dipole moments~\cite{grynberg94,grynberg94a,ackemann95b,leberre95b,scroggie96}. In the meantime, an important research direction established in the atomic, molecular and optical physics community is the simulation of complex quantum systems, e.g.\ from condensated matter physics, using well-controlled systems based on laser cooled or quantum degenerate atomic samples. In particular, the investigation of classical and quantum magnetism is at the center of growing research using ultracold Bose and Fermi gases~\cite{gross17}. Effective exchange interactions have been studied in dilute Bose-Einstein condensates and degenerate Fermi gases at temperatures low enough to observe ferromagnetic or antiferromagnetic coupling~\cite{stenger98,chang04,vengalattore08,kronjager10,jacob12,greif13} including recent progress towards highly configurable simulators~\cite{stuhler05,kaudau16,labuhn16,zeiher17,bernien17}. Hence it appears to be fruitful to look at self-organizing coupled atom-light systems from a complementary point of view with light providing interaction between atoms and not only atoms providing interaction between light waves as in conventional nonlinear optics. Effective long-range interaction between atoms are possible using transversely pumped cavities (see e.g.~\cite{domokos02,black03,baumann10,kollar17,leonard17,landini18}). Furthermore, progress in optical trapping and cooling technology has enabled ensembles with high optical density to make single-mirror feedback and counter-propagating beam experiments feasible in cold atom setups~\cite{greenberg11,gauthier12,labeyrie14,camara15,labeyrie16,schmittberger16,kresic18}. As a first result on magnetic ordering, Ref.~\cite{kresic18} connected magnetic dipole states in the Rb ground state to the transverse Ising model.

However, typically alkaline atoms possess a more complicated ground state not only allowing for magnetic dipole moments (orientation) but also for higher multipoles such as e.g. magnetic quadrupoles (alignments). Contrary to the analogy between magnetic dipole structures and the Ising model~\cite{kresic18}, we are not aware of a specific Hamiltonian describing quadrupole dynamics in condensed matter to be simulated by a diffractive optical feedback scheme but note that recent interest in the community is directed towards investigations in higher-dimensional spaces than offered by spin-1/2 systems, e.g.\ in the spin-1 Haldane model~\cite{senko15,gong16,lee17}. Additional interest stems from the fact that some alignment components represent optically induced coherences between Zeeman sublevels, related to quantum interference phenomena like coherent population trapping, electromagnetically induced transparency and electromagnetically induced absorption~\cite{arimondo76,boller91,akulshin98,fleischhauer05}. There has been a significant interest in exploring spatially structured coherences for image storage for quantum information purposes~\cite{vudyasetu08,shuker08,heinze13,ding13,radwell15}. Some theoretical papers considered the possibility of coherence based self-organized states in cavities~\cite{oppo10,eslami14} but we are not aware of any experiment.

In this paper, we describe the observation of spontaneous spin structures due to light-mediated interaction in a cold thermal cloud. We investigate the rich three-dimensional phase diagram obtained when applying a weak magnetic field to the atoms, and identify phases relying on both dipole and quadrupole moments of the atomic magnetization. In particular, we are reporting on the observation of spontaneous structures in ground state coherences representing magnetic quadrupoles and where they occur in parameter space with respect to the dipole based patterns.

\begin{figure}
\begin{center}
\includegraphics[width=1.0\columnwidth]{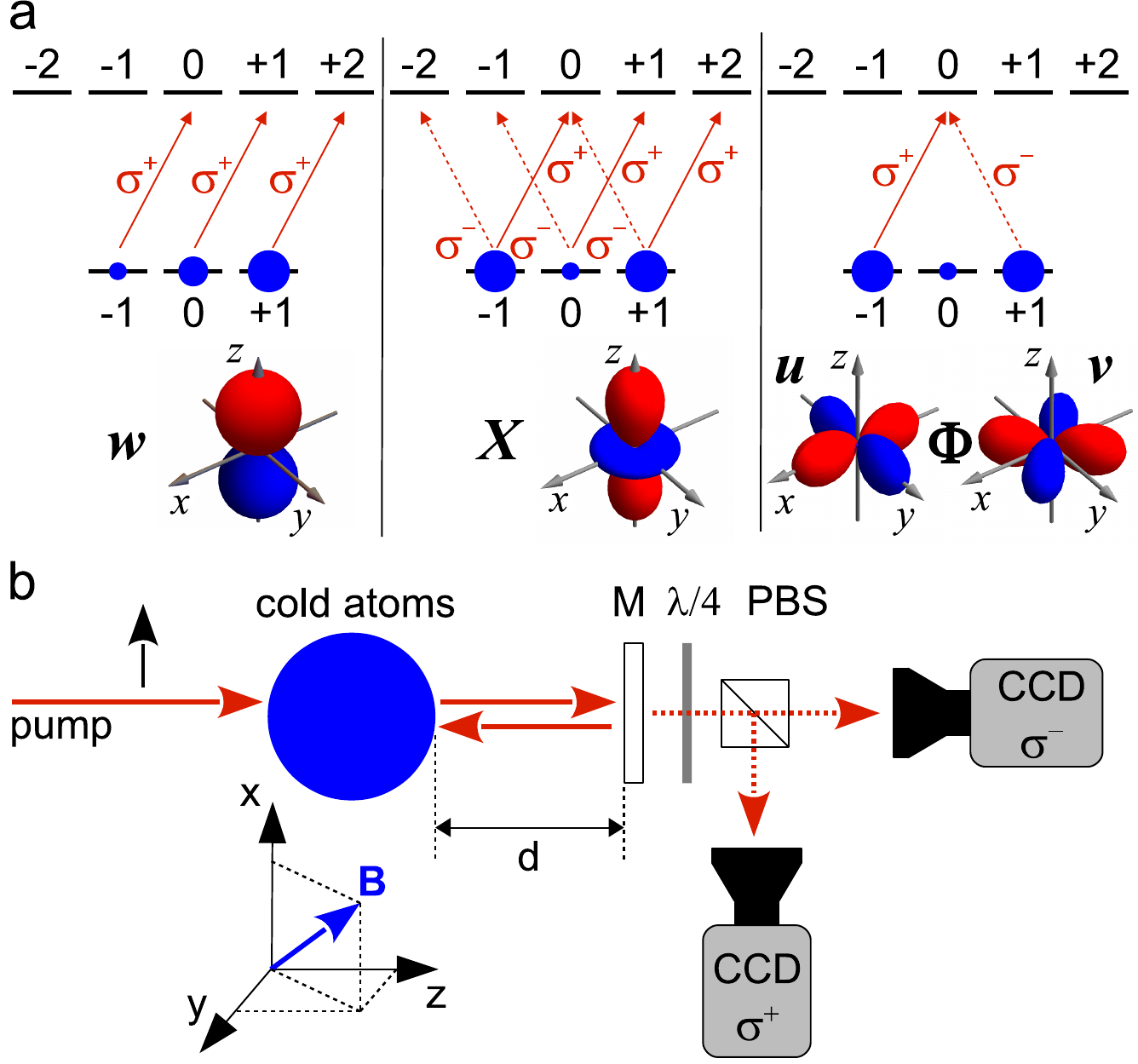}
\caption{Principle of experiment. (a) Components of magnetic moment and symmetries: orientation $w$ (left, dipole), alignment $X$ (middle, quadrupole) and coherence between stretched states $\Phi = u + iv$ (right, quadrupole). (b) Experimental setup. A detuned laser beam is sent through the cold cloud and retro-reflected by a mirror (M). The transverse intensity distribution of the light is recorded in both circular polarization channels, using a quarter-wave plate ($\lambda/4$) + polarizing beam splitter (PBS) assembly.}
\label{fig1}
\end{center}
\end{figure}

The present work builds on earlier experiments on the spontaneous emergence of ordered spatial structures, first in hot vapors~\cite{grynberg94, ackemann94, aumann97}, and then in cold atoms~\cite{gauthier12, labeyrie14, camara15}. As demonstrated in refs.~\cite{grynberg94, ackemann94, aumann97}, the nonlinear interaction between light and atoms can rely on the \textit{magnetic} degrees of freedom. In the following, we describe the atom-light coupling using a $F = 1 \rightarrow F^{\prime} = 2$ transition (see Fig.~\ref{fig1}a). This level structure is simpler than the $F = 2 \rightarrow F^{\prime} = 3$ transition of the $^{87}$Rb D2 line used in the experiment, but still presents components of the magnetic moment beyond the usual dipole. These components, illustrated in Fig.~\ref{fig1}a, are expressed in terms of density matrix elements $\rho_{ij}$. The orientation $w = \rho_{11} - \rho_{-1-1}$ (left) corresponds to an asymmetric population distribution and describes a magnetic dipole oriented along the quantization axis $z$. An orientation can be obtained by Zeeman pumping with circularly-polarized light. The alignment $X = \rho_{11} + \rho_{-1-1} - 2\rho_{00}$ corresponds to a symmetric but uneven population distribution. Whilst $z$ is still the preferred axis, this corresponds to a quadrupole moment~\cite{auzinsh10}. An alignment is obtained through Zeeman pumping with two circular fields with same amplitude but opposite helicities ($\sigma^+$ and $\sigma^-$), corresponding e.g. to a linear polarization orthogonal to $z$. In this situation, the two $\sigma$ fields can also couple to the same excited state as shown on the right of Fig.~\ref{fig1}a, to establish a Zeeman coherence between stretched states $\Phi = 2\rho_{1-1} = u +iv$. $u$ and $v$ correspond to quadrupole states in the (x,y) plane~\cite{auzinsh10}. Both orientation and alignment arise from incoherent processes described by rate equations. In contrast, the description of Zeeman coherences requires the use of optical Bloch equations. For $F^{\prime} < F$ transitions, coherent processes lead to coherent population trapping~\cite{aspect88} and electromagnetically induced transparency~\cite{boller91, radwell15}, and for $F^{\prime} > F$ to electromagnetically induced absorption~\cite{akulshin98}.
In magnetism terminology, the orientation $w$ is proportional to the expectation value of the magnetic dipole operator $F_z$. Alignment $X$ and coherences $u$ and $v$ correspond to magnetic quadrupoles $3F_z^2-F^2$, $F_x^2-F_y^2$ and $F_x F_y + F_y F_x$ respectively~\cite{omont77}.

The dispersive optical properties of the gas are determined by the medium polarization $\mathcal{P}_{\pm}$ induced by the circularly-polarized fields $E_{\pm}$:
\begin{equation} 
\mathcal{P}_{\pm} = \epsilon_0 \mbox{Re}(\chi_{\pm}) \left[ (1 \pm \frac{3}{4} w + \frac{1}{20} X)E_{\pm} + \frac{3}{20}(u \mp iv)E_{\mp} \right]
\label{polarization}
\end{equation}
where $\mbox{Re}(\chi_{\pm}) = \frac{OD}{kL} \frac{2(\delta \mp \Omega_z)/\Gamma}{1 + (2(\delta \mp \Omega_z)/\Gamma)^2}$ is the real part of the linear susceptibility ($k$ is the wavevector, $OD$ the cloud's optical density at resonance, $L$ its thickness, $\delta$ the laser detuning and $\Gamma$ the atomic linewidth). $\Omega_z$ denotes the Zeeman shift in the presence of a longitudinal magnetic field $B_z$. The first term in Eq.~\ref{polarization} describes the phase shift experienced by the $\sigma^{\pm}$ fields, whilst the second one is a conversion term between $\sigma^+$ and $\sigma^-$ in the presence of $\Phi$. The dynamics of $w$, $X$ and $\Phi$ is governed by a set of coupled equations involving light and magnetic field (see Supplementary material~\cite{supplementary}). Thus, the structures discussed here are highly sensitive to both light polarization and magnetic field, in contrast with previous observations in cold atoms~\cite{labeyrie14, camara15}.

The experiment, sketched in Fig.~\ref{fig1}b, is based on the single-mirror feedback setup~\cite{firth90}. A large ($L = 1.4$ cm, OD $\approx 80$, atom number $\approx 10^{11}$) and cold ($\approx 200~\mu$K) atomic cloud released from a magneto-optical trap is illuminated by a pulsed ''pump'' laser beam (pulse duration 400$~\mu$s) of waist 2.2 mm and peak intensity $I_0$ propagating along $z$ and linearly-polarized along $x$. This laser is detuned from the atomic transition such that single-pass absorption is moderate (typically 20$\%$), justifying the dispersive description of Eq.~\ref{polarization}. The transmitted beam is retro-reflected by a semi-transparent mirror ($R > 99\%$) located at a distance $d$ behind the cloud. A typical mechanism for self-organization is as follows. Consider a local, microscopic fluctuation of the orientation $w(x,y)$ resulting in a phase difference between transmitted $\sigma^+$ and $\sigma^-$ fields. After diffractive propagation over $2d$, this phase difference turns into an intensity imbalance yielding a differential Zeeman pumping which enhances the initial orientation fluctuation. Above a certain intensity threshold, spontaneous symmetry breaking occurs leading to the formation of two-dimensional spatial structures in the transverse plane for both light and orientation. The light transmitted by the mirror is used to image the transverse intensity distribution of the beam either in near-field (NF) or in far-field (FF). This imaging can be performed simultaneously in two orthogonal polarization channels, either circular ($\sigma^+$/$\sigma^-$) or linear (termed $//$ and $\perp$ for a polarization aligned with that of the pump, or orthogonal to it).

\begin{figure*}[t!]
\begin{center}
\includegraphics[width=2\columnwidth]{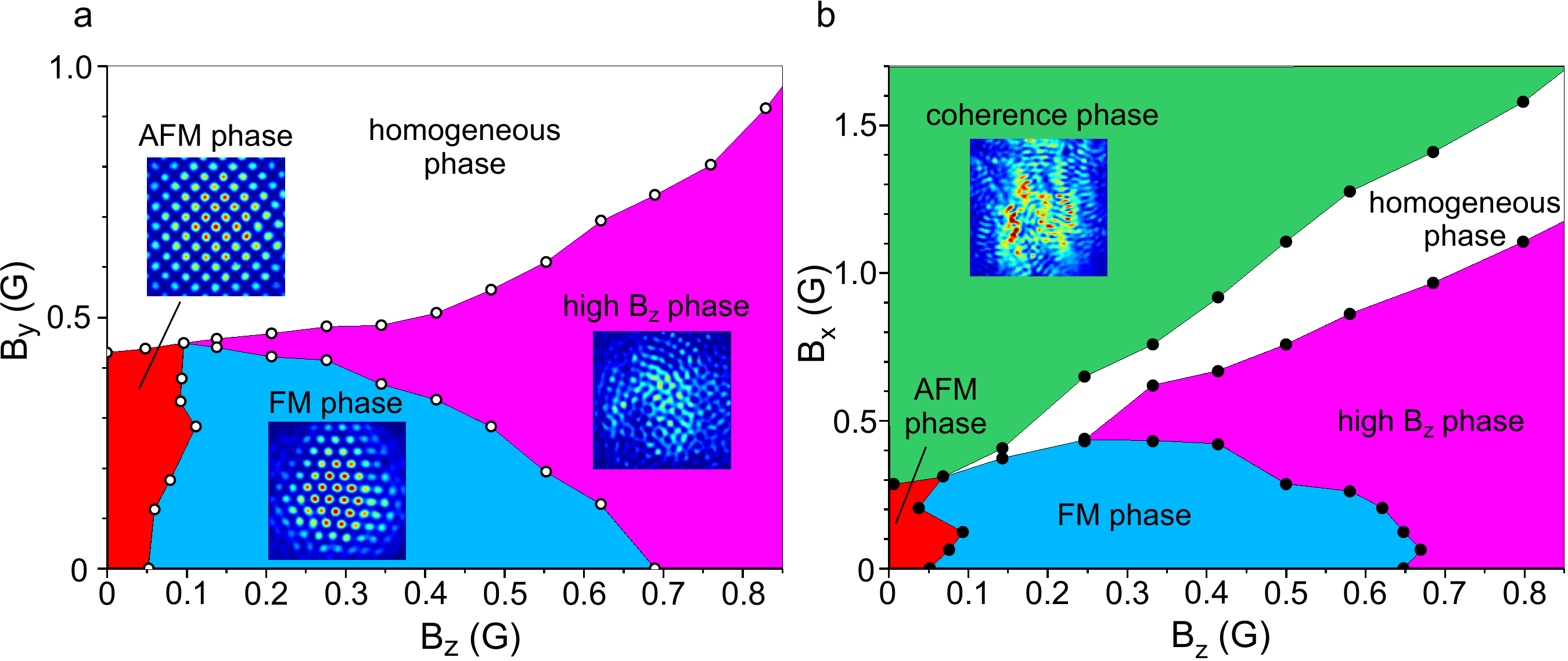}
\caption{Magnetic phase space of spin structures (experimental). (a) $B_y-B_z$ cross-section of phase space. Three phases with different symmetries are observed as illustrated by NF images. These phases vanish for a large transverse field $B_y$. (b) $B_x-B_z$ cross-section of phase space. An additional phase is observed when the transverse field $B_x$ is increased. The structures typically take a few $10-100~\mu$s to emerge, and can persist for a few ms. Parameters: $OD = 80$, $I_0 = 8$ mW/cm$^2$, $\delta = -8\Gamma$, $d = -20$ mm.}
\label{fig2}
\end{center}
\end{figure*}

An intriguing feature of cold atomic samples is the simultaneous presence of several mechanisms leading to the formation of spatial structures~\cite{labeyrie14, camara15}. We can select a specific mechanism by choosing the appropriate experimental parameters. To observe spin structures, we set the laser detuning at a negative value (typically $\delta = - 8\Gamma$) since opto-mechanical~\cite{labeyrie14} and saturation~\cite{camara15} patterns are only observed for $\delta > 0$. We use a laser intensity as low as 1 mW/cm$^2$, typically two orders of magnitude below the thresholds of these non-magnetic instabilities. The excited state population is thus low ($\approx 10^{-3}$) and pattern formation is dominated by ground-state physics. We also cancel the residual magnetic field down to $\approx 10$ mG in all three dimensions.

We then apply a control magnetic field, and vary its direction and magnitude to observe the complex phase space reported in Fig.~\ref{fig2}. Each phase is characterized by a specific spatial distribution of the light intensity and of the underlying atomic spins. We observe a phase with square symmetry (AFM, in red) localized around $B = 0$. Slightly increasing  the longitudinal magnetic field $B_z$ leads to an hexagonal phase (FM, in blue). A larger $B_z$ eventually produces a phase without long range order, but with a local remaining hexagonal symmetry (''high $B_z$'', in magenta). All these phases vanish when the transverse field $B_y$ is increased (Fig.~\ref{fig2}a). On the contrary, a new disordered phase with a peculiar symmetry (''coherence'', in green) appears when increasing $B_x$ (Fig.~\ref{fig2}b).

This phase space depends on experimental parameters such as OD and laser intensity. Reducing the OD leads to the appearance of gaps between the AFM and coherence phases on one hand, and between the FM and high $B_z$ phases on the other hand. Lowering the OD to 40 lead to the vanishing of the coherence phase. This proves that these phases are of a different nature, with different OD thresholds. Increasing $I_0$ leads to a broadening of all the features seen in Fig.~\ref{fig2}, because the boundary for a given phase is typically determined by the balance between a Larmor and a Rabi frequency. For instance, Zeeman pumping is hampered by a transverse magnetic field inducing coherent coupling between Zeeman substates~\cite{kresic18}. Increasing $I_0$ requires a proportional increase of the magnetic field to obtain the same steady-state population.

\begin{figure}
\begin{center}
\includegraphics[width=1.0\columnwidth]{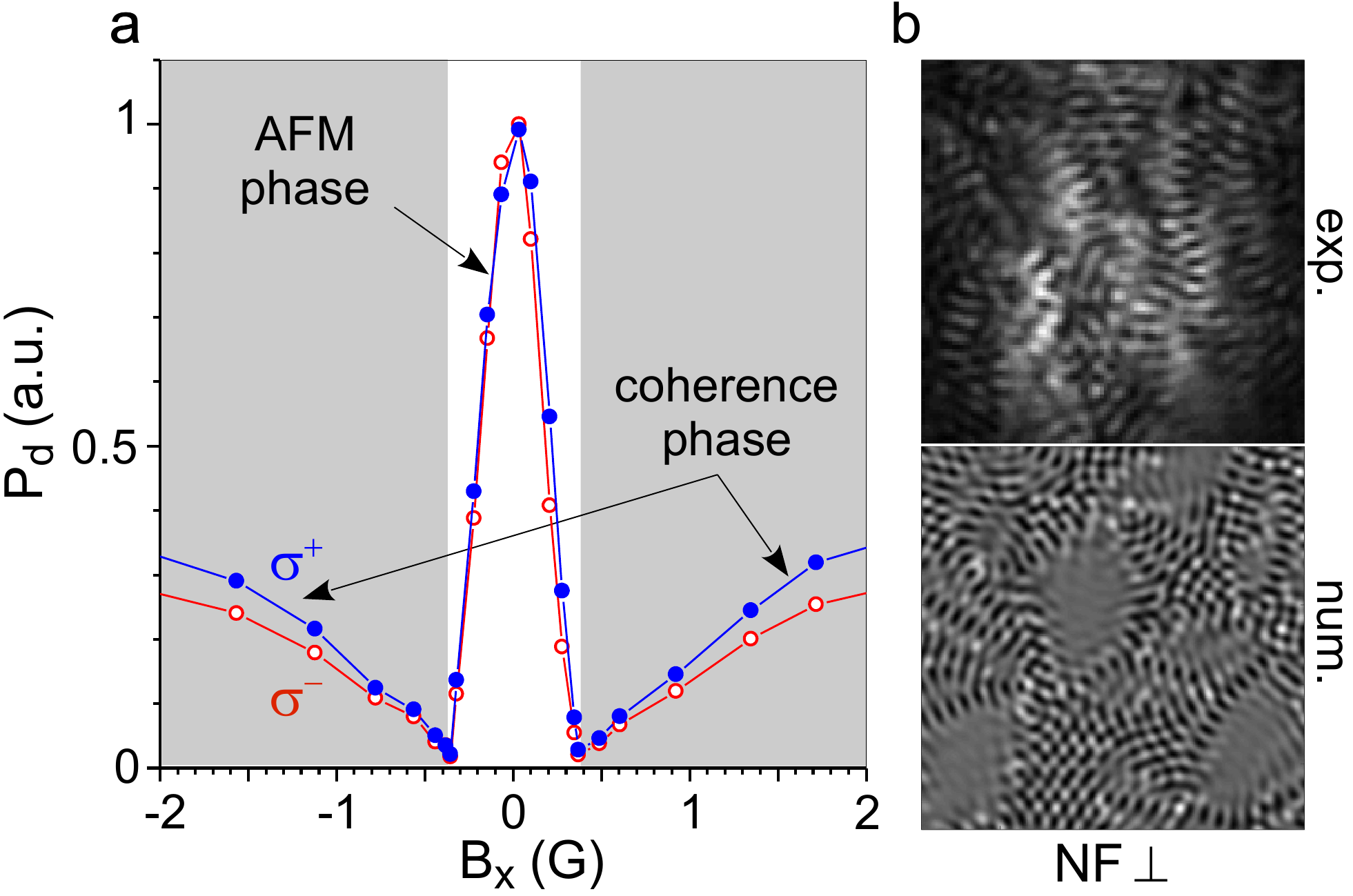}
\caption{Coherence phase. (a) $B_x$ scan showing the transition from the AFM phase (white background) to the coherence phase (shaded). Parameters: $B_y = B_z = 0$, $I_0 = 12$ mW/cm$^2$, $\delta = -8\Gamma$, $d = -20$ mm. (b) Experimental (top) and numerical (bottom) NF patterns for $B_x = 1$ G, both in the $\perp$ channel. The field of view of the experimental image is 3.3 mm, the typical pattern length scale is 170 $\mu$m.}
\label{fig3}
\end{center}
\end{figure}

Orientation has the highest prefactor in Eq.~\ref{polarization} and is thus expected to dominate the formation of spin structures, at least for weak transverse magnetic fields. Indeed, a detailed theoretical analysis beyond the scope of this paper reveals that the sequence AFM $\rightarrow$ FM $\rightarrow$ high-$B_z$ phase observed when increasing $B_z$ can be understood considering only the $w$ term in Eq.~\ref{polarization}. However, the coupling of $w$ to the other atomic quantities~\cite{supplementary} needs to be retained in the analysis. We have also shown that the AFM and FM phases correspond respectively to anti-ferromagnetic and ferrimagnetic arrangements of atomic spin domains interacting via light. These findings as well as analogies with the quantum Ising model are discussed elsewhere~\cite{kresic18}.

We concentrate in the following on original spin phases that require the presence of quadrupole terms in the magnetic moment. This interesting physics extending beyond spin-1/2 models arises from our extended Zeeman structure. To promote these phases, we need to overcome the low coupling efficiencies associated with the $X$, $u$ and $v$ terms in Eq.~\ref{polarization}. This can be achieved by two means: increasing the transverse magnetic field to destroy orientation, or manipulating the effective interaction between spins by modifying the polarization of the feedback light.

We illustrate the first case in Fig.~\ref{fig3}a, where we plot the ''diffracted power'' $P_d$ corresponding to the total power of the spatially modulated light extracted from the FF images as a function of $B_x$ ($B_{y,z} = 0$). $P_d$ is recorded in both circular channels. The maximum at $B_x = 0$ corresponds to the orientation-based AFM phase of Fig.~\ref{fig2}. Increasing $\left|B_x\right|$ leads to a rapid decrease of $P_d$ and to the disappearance of the AFM phase around $\left|B_x\right| = 0.37$ G. When $\left|B_x\right|$ is further increased, the coherence phase appears (shaded area). The amplitudes of the $\sigma^+$ and $\sigma^-$ fields remain approximately equal, which is indeed expected for $B_z = 0$ as the Zeeman structure is symmetric (Fig.~\ref{fig1}a). We observe that the difference $\sigma^+ - \sigma^-$, which drives the growth of the orientation~\cite{supplementary}, doesn't show any spatial modulation in the coherence phase. It is on the contrary strongly modulated for the AFM, FM and high $B_z$ phases. For both phases of Fig.~\ref{fig3}a, most of the spatially-modulated light is generated in the $\perp$ channel (polarization instability). This can be understood by looking at Eq.~\ref{polarization}. The terms with $\pm$ or $\mp$ signs in front ($w$ and $v$) yield a different optical response for $\sigma^+$ and $\sigma^-$ fields, and can thus change the polarization. On the opposite, $X$ and $u$ terms yield the same response to $\sigma^+$ and $\sigma^-$ and hence do not modify the polarization.

\begin{figure}
\begin{center}
\includegraphics[width=1.0\columnwidth]{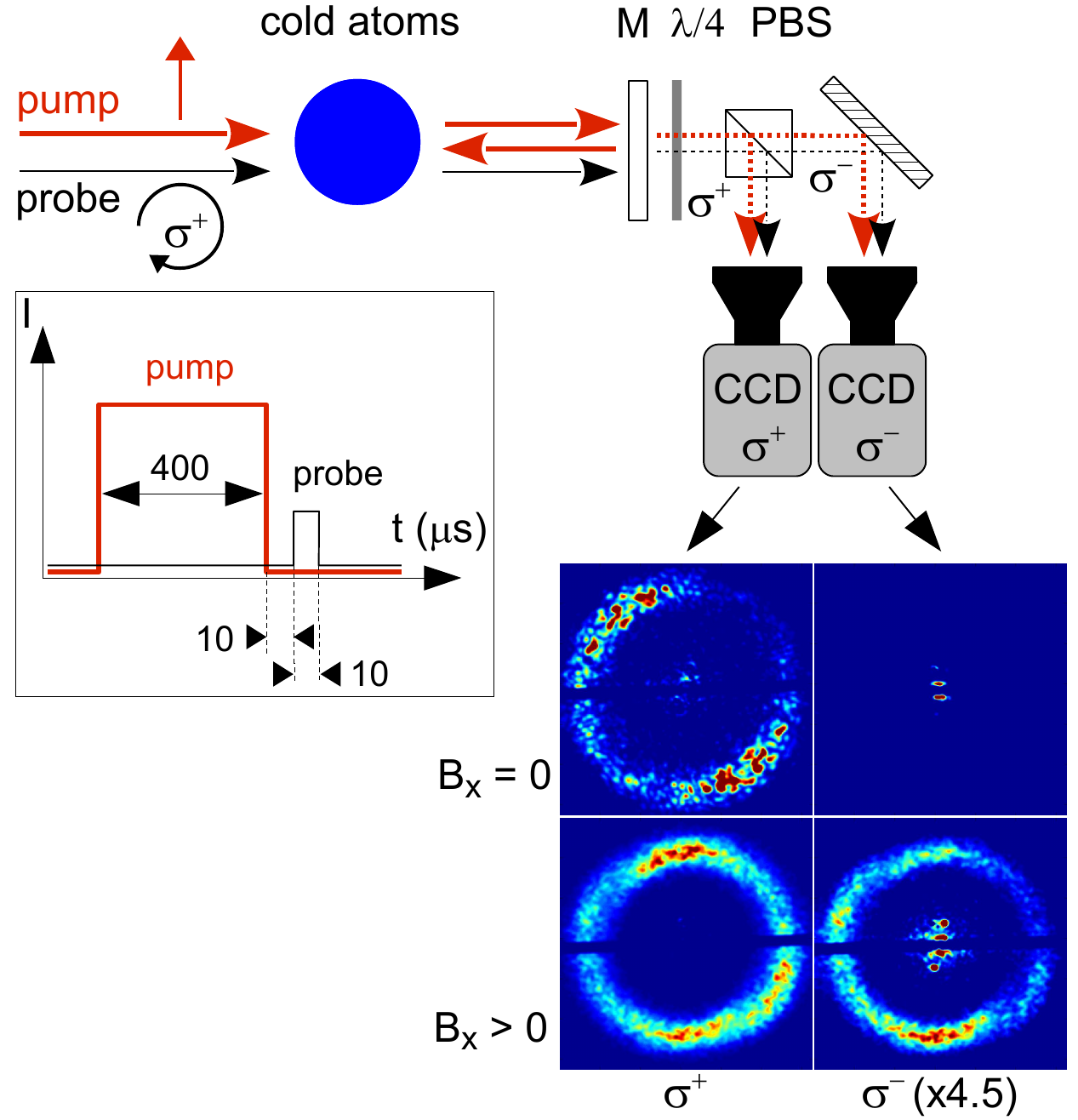}
\caption{Experimental evidence for a strong spatial modulation of $\Phi$ in the coherence phase. A $\sigma^+$ weak probe is sent through the cloud shortly after the pump, and FF images are detected in the $\sigma^+$ and $\sigma^-$ channels. For $B_x = 0$, all the spatially-modulated probe light is detected in $\sigma^+$. On the contrary, for $B_x = 1.5$ G, approximately $20 \%$ of the spatially-modulated probe light is transferred to the $\sigma^-$ channel.}
\label{fig4}
\end{center}
\end{figure}

These observations suggest that $v$ is responsible for the coherence phase. To confirm this hypothesis, we performed the experimental test illustrated in Fig.~\ref{fig4}. Immediately after the pump pulse, we sent a much weaker probe pulse of $\sigma^+$ polarization. If a spatially-modulated coherence is present and provides a ''cross-channel'' gain between $\sigma^+$ and $\sigma^-$, we expect some sizable amount of light to be transferred to the $\sigma^-$ channel. This was indeed observed for $B_x = 1.5$ G, with up to $20~\%$ of spatially-modulated probe light detected in the $\sigma^-$ channel. For $B_x = 0$, the detected amount was negligible. This observation proves that a large spatial modulation of $\Phi$ is present when the structures shown in Fig.~\ref{fig3}b are observed. Our numerical simulations indeed confirm the existence of an instability at large $B_x$, associated with a spatial modulation of $v$ one order of magnitude larger than observed for $B = 0$. The numerical patterns displayed in Fig.~\ref{fig3}b (bottom) are qualitatively rather similar to those observed in the experiment (Fig.~\ref{fig3}b, top): in particular, the peculiar arrangement of these structures with patches, sinuous lines of defects and a lack of clear global symmetry is well reproduced. We stress that a quantitative agreement is beyond our present simulation capabilities, both because of our simplified transition model and the fact that thick-medium effects~\cite{firth17} are neglected. The simulations however allow direct access to the atomic states~\cite{supplementary}. Our analysis shows that the linearly polarized pump beam only creates the $u$ component of the coherence, which corresponds to a quadrupole aligned along the x- and y-axes (Fig. 1a, right column). The instability creates a $v$-component with a spatially-modulated amplitude (the amplitude of $u$ being modulated as well). The result is a state with a quadrupole moment of spatially-modulated amplitude, and whose axes are oscillating in space around the x and y axes. In optical terms, this corresponds to a coherence $\Phi$ whose amplitude and phase are spatially modulated.

\begin{figure}
\begin{center}
\includegraphics[width=1.0\columnwidth]{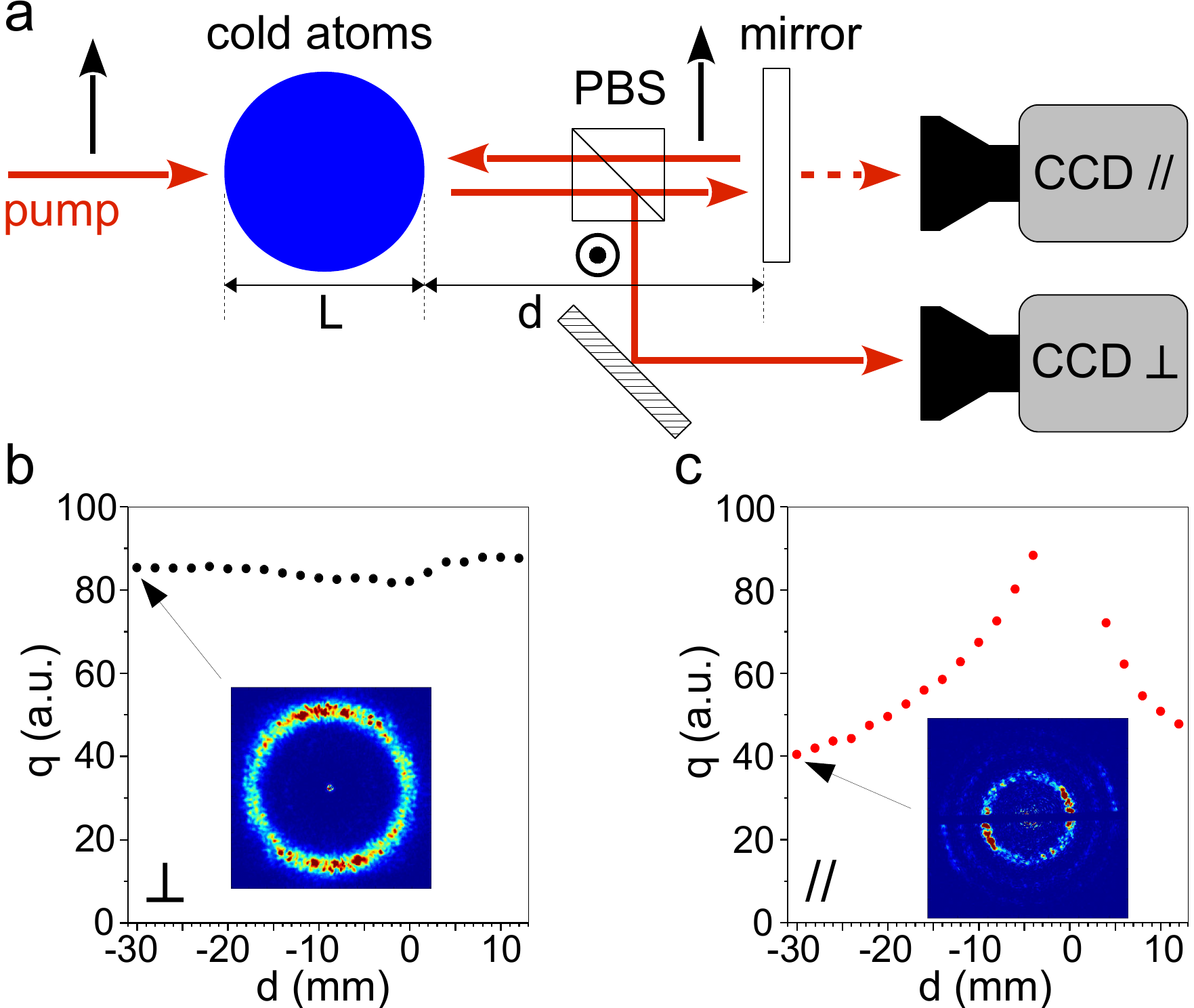}
\caption{Alignment phase. (a) Modified setup with a PBS inside the feedback loop. We monitor the wavevector $q$ of the spatial structures versus feedback distance $d$ in both // and $\perp$ polarization channels. (b) $q$ vs $d$ in the $\perp$ channel. $q$ is nearly $d$-independent (no mirror feedback). (c) $q$ vs $d$ in the // channel. Active mirror feedback hints at a polarization-preserving instability mechanism. $B = 0$, $I_0 = 53$ mW/cm$^2$, $\delta = -8\Gamma$.}
\label{fig5}
\end{center}
\end{figure}

As mentioned above, the phases observed around $B=0$ are essentially orientation-based. In principle, the structure of our ground state can also support a modulated alignment $X(x, y)$, with an associated novel phase. An $X$-based instability is however not favored according to Eq.~\ref{polarization} because of its small prefactor. As alignment is as sensitive to transverse $B$ fields as orientation, we need to selectively suppress the orientation mechanism at $B = 0$. To this end, we inserted inside the feedback loop a PBS aligned to transmit the pump polarization (Fig.~\ref{fig5}a). By imposing a polarization-maintaining feedback, we aim at suppressing the $w$-based instability (and any polarization instability for that matter). On the contrary, we expect a $X$-instability to remain unaffected since the alignment term cannot sustain a polarization instability. We detect the spatially-modulated light both in the // channel (transmitted by the PBS and the mirror) and in the $\perp$ channel (now rejected by the PBS). The FF images in Fig.~\ref{fig5}b and c show that spatial structures are observed in both polarization channels. In the plots of Fig.~\ref{fig5}b and c, we record the transverse wave vector $q$ of the diffracted light as a function of the feedback distance $d$. Negative values of $d$ are accessible through the use of an imaging system~\cite{ciaramella93}, which is not shown. A variation of $q$ with $d$ is a signature of mirror feedback~\cite{firth90}. We observe that in the // channel (Fig.~\ref{fig5}c) the mirror feedback is still operational: $q$ varies with $d$. A polarization-preserving mechanism is thus at work, which from the previous discussion can only involve $X$ or $u$. We observed that the intensity threshold of this instability is higher by a factor $3$ than that of the $w$-instability (in the absence of PBS inside the feedback loop). The structures vanish if we increase the transverse magnetic field $B_x$. These observations appear to be consistent with a role played by the alignment. A mechanism based on the $u$ term seems unlikely, since the role of the coherence $\Phi$ is favored by $B_x$ (Fig~\ref{fig3}). Surprisingly, we also observe patterns in the $\perp$ channel, but with a $d$-\textit{independent} wave vector(Fig.~\ref{fig5}b). Indeed, due to the large thickness of our atomic cloud, diffraction takes place inside the cloud and we can observe instabilities using two ''independent'' counter-propagating beams instead of a retro-reflected one~\cite{labeyrie16}. In this situation, the wave vector of the instability is determined by the cloud's thickness $L$ and is independent of $d$. We thus conjecture that two instabilities coexist in the presence of the PBS inside the feedback loop : an alignment-based instability (without polarization instability) with mirror feedback in the // channel, and an orientation-based instability (with polarization instability) without mirror feedback in the $\perp$ channel. This unprecedented and complicated situation obviously requires more investigations, since thick-medium effects~\cite{firth17} have to be included in the theoretical analysis.

In conclusion, we investigated the spontaneous formation of spatial spin structures in a cold atomic cloud submitted to optical feedback. This system, coherently-driven and dissipative~\cite{kessler12}, displays an unconventional form of magnetism where atomic spins interact non-locally via light. It offers the possibility to tailor effective spin-spin interactions through modifications of the feedback loop. By tuning a weak magnetic field, we induced transitions between phases with various symmetries relying on the spatial modulation of different components of the atomic magnetic moment, corresponding to both dipole and quadrupole terms. In particular, we observed an original phase based on ground-state Zeeman coherences which was not accessible in previous spin-1/2 studies. This could spark a renewed interest in the quest to store spatial information in atomic coherences for quantum memories~\cite{vudyasetu08,shuker08,heinze13,ding13,radwell15}. Future directions include the investigation of other atomic transitions, the statistics of symmetry breaking and Kibble-Zurek dynamics~\cite{labeyrie16}, the influence of frustration, and the search for optically controllable localized magnetic structures.

\noindent
{\large \bf Acknowledgements}\\
{\footnotesize The Strathclyde group is grateful for support from the Leverhulme Trust. IK gratefully acknowledges a University Studentship from the University of Strathclyde. The Nice group acknowledges support from CNRS, UNS, and R\'{e}gion PACA. The collaboration between the two groups was supported by the Royal Society (London), CNRS and in particular the Laboratoire International Associ\'{e} ''Solace'', and the Global Engagement Fund of the University of Strathclyde. The final stage of this work was performed in the framework of the European Training network ColOpt, which is funded by the European Union Horizon 2020 program under the Marie Skodowska-Curie action, grant agreement 721465.

\end{document}